# Resonance and Depth: Composing Timbral Identity and Spatial Form in an Artist-in-the-Loop Co-Creative Multi-Agent Percussive Performance

Keon Ju Maverick Lee
School of Interactive Arts and Technology, Simon Fraser University, Vancouver, British Columbia, Canada

Corresponding author: keon_maverick@sfu.ca

This research-creation [1] artistic report presents *Resonance and Depth*, a 9-minute immersive electroacoustic/acousmatic [2] performance that treats timbre and space as primary compositional materials. Built from live electronic percussion and a personal corpus recorded, curated, and composed by the author, including metals, skins, frictional noise, resonant objects, and chromatic percussive materials, the work investigates how timbre-as-structure can emerge through real-time recombination rather than conventional harmonic development. Rather than positioning rhythm, pitch, or harmony as the main carriers of form, *Resonance and Depth* foregrounds timbral morphology, spatial movement, and performer–system interaction.

Situated within a research-creation methodology [1], the work develops artistic practice, technical experimentation, and theoretical reflection together. The performance is not simply a demonstration of a technological system, but an inquiry into how performer identity can be retained, transformed, and redistributed through co-creative machine processes. The personal corpus functions as both musical memory and compositional material: it carries traces of the performer's instrumental vocabulary while becoming available for algorithmic recombination, fragmentation, and spatial reconfiguration.

The piece is realised as a musical multi-agent society/system [3,4] designed for artist-in-the-loop music improvisation in a co-creative space [5]. At the macro scale, an evolving generative ambient layer shapes atmosphere, density, and long-form spectral drift. At the micro scale, the performer's rhythmic gestures—short transients, rolls, frictional attacks, and resonance tails—drive a call-and-response relationship with the agents, producing tension–release passages across the performance. The agents reshape the performer's material into evolving textures, while the performer responds to the changing timbral field, timing, density, and spatial behaviour.

The performance workflow is coordinated by a conductor agent that supports structured improvisation across the piece. It monitors scene-level changes such as density, energy, texture, musical tension, and agent states, then automates behavioural shifts across the system. This enables musical acts to emerge dynamically while preserving an overall formal arc. The workflow can be summarised as follows:

- **Human performer / Drumbeam–multipad percussion interface:** Provides embodied rhythmic input through electronic percussion while extending MIDI control from discrete note triggering to continuous gestural control of timbre, morphology, and instrumental identity.
- **MACAT:** Recombines the author's personal corpus as a timbral memory space.
- **MACataRT:** Generates audio-mosaicing responses from the live stream and corpus.
- **MIDI-based agent:** Translates performance activity into symbolic output using audio-to-MIDI transcription for machine listening and Karplus–Strong sound synthesis.
- **Spatialisation layer:** Distributes sound objects through trajectory, diffusion, depth, width, and apparent-size controls.

- **Conductor agent:** Coordinates scene changes and automates agent behaviours for the structured music improvisation, enabling transitions between sparse ambience, dense call-and-response, close mirroring, and offers more autonomy for the human performer during the performance.

Agent 1, MACAT [6], acts as the primary timbral recombination engine. It traverses a personal corpus as a timbre space, sequencing fragments into emergent phrases that remain connected to the performer's instrumental identity while being algorithmically recontextualised. A synthesis layer, including resampling, reversal, pitch transformations, and related sound-design processes, further reshapes retrieved materials into malleable sound objects, sustaining a continuity between instrumental causality and acousmatic ambiguity.

Agent 2, MACataRT [6], introduces a musique-concrète counter-voice through audio mosaicing [7] for creating reactive music accompaniment. It listens to the live stream and reassembles corpus fragments by feature similarity, producing doubles, shadows, echoes, and swarms around the performer and another agent's audio inputs. This agent can move from close mirroring to more autonomous textural divergence, creating an intermediate space between accompaniment and autonomous agency.

Agent 3, a MIDI-based agent presented in the musical agents tutorial [8], listens to the audio agents through audio-to-MIDI transcription and/or direct MIDI input. By translating audio-agent activity and live percussive input into symbolic musical information, this agent introduces a MIDI-domain layer of mediation between sound, gesture, and system response. Its MIDI output is rendered through MIDI-based percussion and a Karplus–Strong string synthesiser [9], producing a distinctive timbral palette that contrasts with, while also resonating against, the metallic, skin-based, and frictional materials of the original corpus. This symbolic-to-synthetic layer explores how percussive and acousmatic materials can be reinterpreted through symbolic control, physically inspired synthesis, and MIDI-driven timbral articulation, a topic that remains relatively underexplored compared with audio-domain approaches.

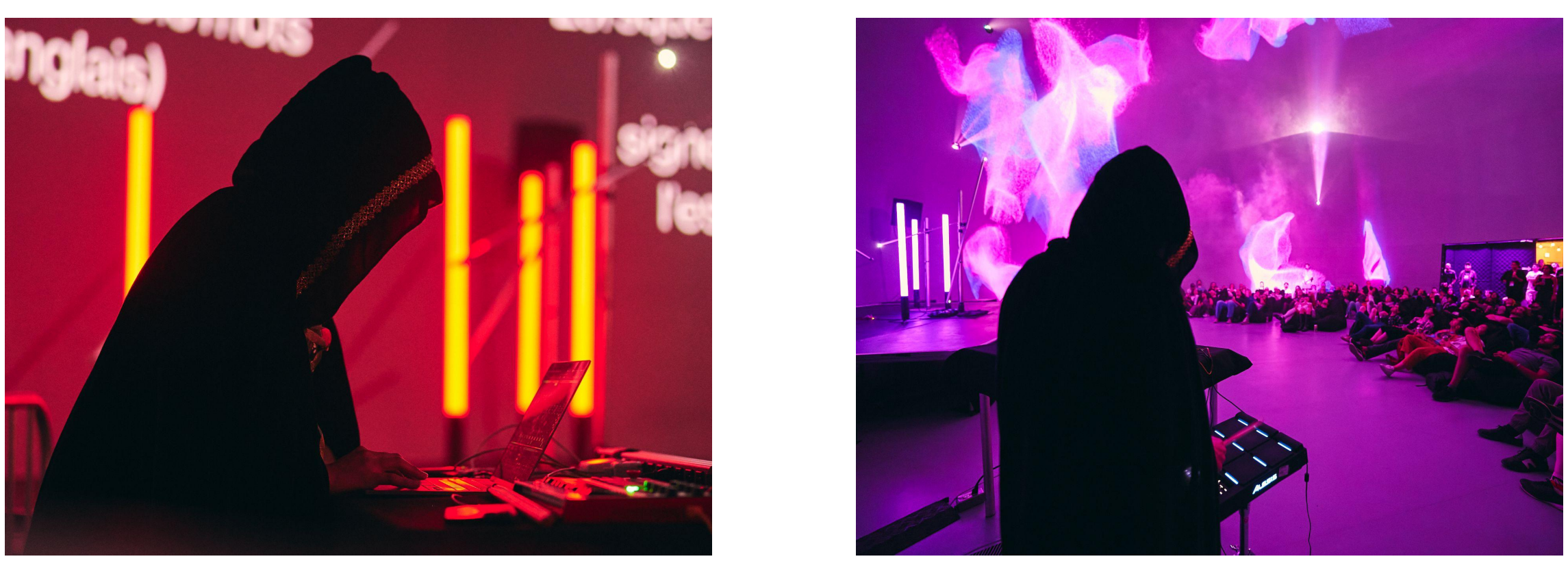

*Figure 1: Documentation of 93.5 mix immersive dome performance at the société des arts technologiques during the MUTEK Montréal Festival, 2025. Photo credit: Bruno Aiello-Destombes.*

This MIDI-based layer also reframes spatial agency in relation to recent audio-domain work on spatial awareness and influence in co-creative agents. Gatti and Fiorini [10] propose an automated corpus-spatialisation system for Somax2 in which audio descriptors, transient behaviour, and agent roles are mapped to spatial trajectories and inter-agent influence. In *Resonance and Depth*, a related idea is translated into the MIDI domain: notes, velocity, timing, CC/MPE data, and percussive control streams operate not only as triggers, but also as spatial–timbral control signals. This extends spatial influence from audio-feature-driven spatialisation toward symbolic, performative control, where MIDI becomes a medium for composing timbral identity and spatial form.

Spatialisation is treated as composition rather than post-mix, following Normandeau's argument that space can function as a primary musical parameter in acousmatic music and that timbre itself can be distributed across loudspeakers as "timbre spatialisation" [11]. This approach extends the author's earlier sound spatialisation practice developed during an artist residency at the Société des arts technologiques (SAT), where immersive diffusion and spatial motion were explored as compositional materials within dome-based immersive auditory environments. In *Resonance and Depth*, this framework is expanded through musical agents and an object-based 3D approach [12], where each sound source is authored with evolving position, trajectory, depth, width, apparent size, and diffusion. The original corpus is designed for the 93.5-loudspeaker dome, where stochastic motion and diffusion trajectories act as timbral gestures: distance, orbit, elevation, and spatial density become parameters of phrasing and form.

Here, spatialisation is part of a dynamic co-creative ecology in which human performance, generative systems, concatenative processes, and spatial behaviours influence one another in real-time. The incorporation of musical agents transforms the immersive environment into an adaptive performance ecosystem where timbral recombination, mosaicing, and spatial trajectories become active participants in the musical discourse. For portability, the same object scene can be re-imagined across formats; this performance version is provided as stereo/binaural documentation, with an optional 5.1 theatre art mix or spatial dome mix if an audio system is available.

The Drumbeam instrument integrated with a percussive multipad supports the exploration of timbral identity in the MIDI domain by extending percussive gesture beyond discrete note triggering into continuous, multidimensional control. Parameters such as strike intensity, spatial position, and gestural variation can be mapped to MIDI CC or MPE data, allowing the performer to shape timbre as an expressive continuum. Within the broader system, rhythmic performance becomes a practice of timbral articulation, morphological variation, and instrumental redefinition within symbolic control systems.

For the Timbre conference, the work contributes three main perspectives. First, it proposes a practice-based method for crafting authored corpora that retain performer identity within concatenative sound systems [13]. Second, it presents a model of co-creative timbral agency in which recombination, mosaicing, and symbolic mediation act as improvisational partners rather than as purely automatic generation processes. Third, it offers an immersive case study of timbre spatialisation, showing how psychoacoustic attributes can be composed alongside instrumental gesture to produce musical form. Through these contributions, *Resonance and Depth* frames timbre as a generative, spatial, and relational force through which performer, machine, and listening environment co-produce the work.

**Resonance and Depth (2026) — Selected Listening / Bandcamp Album Release**

Live spatial performance presented during an artist residency at the Société des arts technologiques (SAT) in Montréal, an internationally recognized centre for immersive and digital arts known for its Satosphere multichannel dome, during MUTEK Montréal, an international festival dedicated to electronic music, digital creativity, and audiovisual experimentation. The work formed part of an approximately 30-minute live set presented in a 93.5-channel spatial mix. The continued development of the project is supported by the Glenfraser Acoustic Communication Research Award, endorsed by Prof. Barry Truax and Dr. Guenther Krueger.

**Album release for listening:**

https://keonlanocturneartist.bandcamp.com/album/hyperspatial-drum-ritual